# Lowering the lasing threshold of distributed feedback lasers with loss

V. Yu. Shishkov,[1,2] A. A. Zyablovsky,[1,2] E. S. Andrianov,[1,2] A. A. Pukhov,[1,2,3] A. P. Vinogradov,[1,2,3] A. V. Dorofeenko[1,2,3] and A.A. Lisyansky[4,5]

[1]*Moscow Institute of Physics and Technology, 9 Institutskiy per., 141700 Dolgoprudny, Moscow reg., Russia*
[2]*All-Russia Research Institute of Automatics, 22 Sushchevskaya, Moscow 127055, Russia*
[3]*Institute for Theoretical and Applied Electromagnetics RAS, 13 Izhorskaya, Moscow 125412, Russia*
[4]*Department of Physics, Queens College of the City University of New York, Flushing, NY 11367, USA*
[5]*The Graduate Center of the City University of New York, New York, New York 10016, USA*

We study laser generation in 1D distributed feedback lasers with amplifying and lossy layers. We show that when the lasing frequency differs from the transition frequencies of the amplifying medium, loss induced lasing may occur due to the broadening of the resonator mode with increasing loss in the absorbing layers. This broadening leads to a shift in the lasing frequency towards the transition frequency. As a result, the cavity mode interaction with the amplifying medium is enhanced, and the lasing threshold is lowered.

## I. INTRODUCTION

Lasers generate coherent radiation due to induced transitions in an active medium caused by the electromagnetic field in the resonator. For laser generation to arise, in addition to the resonant phase conditions being fulfilled, the amplification of the field must be sufficient to compensate for loss due to dissipation in the laser material and radiation from the sample.[1, 2] It is, therefore, natural to expect that an increasing resonator loss would result in an increase of the lasing threshold. There are situations, however, when the lasing threshold decreases as loss increases.[3, 4]

In Ref. 3, quenching of lasing by increasing the strength of the pump has been predicted for a system of two identical resonators containing an amplifying medium kept at different pump rates. This results in the field concentration in the more strongly pumped resonator. Lasing starts in this resonator. Due to coupling of resonator modes, the oscillations in the second resonator are synchronized to the lasing mode and the whole system lases. When the pump rate in the low pumped resonator increases while remaining fixed at the other one, a phase transition from a non-symmetric to symmetric field distribution in the resonators occurs. As a result, the field increases in the low pumped resonator and decreases in the highly pumped resonator leading to the suppression of lasing. In experiment,[4] instead of increasing the pump rate, the loss in one of the resonators was increased, and the phase transition went from symmetric to the non-symmetric eigenmode. This ensured that the optical field is concentrated in the more active resonator and lowered the lasing threshold.

Another possibility of the lasing onset may be realized when the increase in the loss causes a variation of the refractive index and improves the phase condition. This was shown theoretically for a resonator uniformly filled with an active medium with temporal dispersion.[5]

In lasers, the transition frequency $\omega_0$ of the amplifying medium is usually tuned to the frequency of one of the resonator modes $\omega_R$. This allows for the greatest effective interaction between the field and the amplifying medium. In real systems, however, these two frequencies are detuned. Among other reasons, a detuning between $\omega_0$ and $\omega_R$, which depend differently on the temperature, may arise due to the temperature variation in the system during the process of lasing.[6]

In this paper, we demonstrate a mechanism of lasing generation via loss that works when the frequency of the transition line of the amplifying medium differs from the laser mode frequency. An increase of loss may pull the lasing frequency towards the transition frequency thereby lowering the laser threshold. Analytical calculations are confirmed by computer simulation for a distributed feedback (DFB) lasing possessing both amplifying and absorbing layers. The obtained results can be used for lowering laser generation thresholds in DFB lasers with metallic layers, which have recently been actively studied.[7-13]

## II.   MAIN EQUATIONS

We consider a DFB laser based on a one-dimensional photonic crystal (see Fig. 1)[13]. The elementary cell of this crystal consists of passive metallic and active dielectric layers. As we show below, thanks to the metal layers, the field distribution in the system changes only slightly when loss increases. As a result, the lasing mode does not shift when the loss level changes.

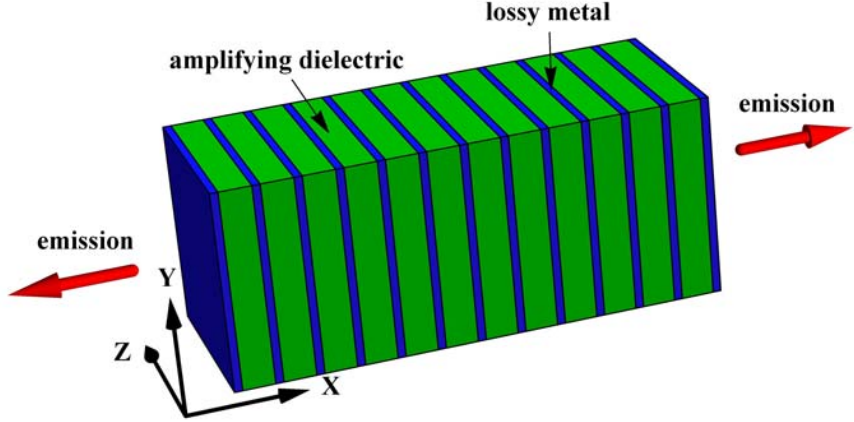

FIG. 1. Schematic of the DFB laser based on 1D photonic crystal. Thin blue and thick green layers depict passive metallic and active dielectric layers, respectively.

The interaction of the electromagnetic field $E$ with the active medium can be described with the Maxwell-Bloch equations[2, 14]

$$\frac{\partial^2 E}{\partial x^2} - \frac{\varepsilon(x)}{c^2}\frac{\partial^2 E}{\partial t^2} = \frac{4\pi}{c^2}\frac{\partial^2 P}{\partial t^2}, \qquad (1)$$

$$\frac{\partial^2 P}{\partial t^2} + \frac{2}{\tau_P}\frac{\partial P}{\partial t} + \omega_0^2 P = -\frac{2\omega_0 |\mathbf{d}_{ge}|^2 nE}{\hbar}, \qquad (2)$$

$$\frac{\partial n}{\partial t} + \frac{1}{\tau_n}(n - n_0) = \frac{2E}{\hbar\omega_0}\frac{\partial P}{\partial t}, \qquad (3)$$

where $P$ is the polarization of the active medium which we consider as a two-level system (TLS), $n$ is the population inversion of the TLS, $\omega_0$ and the transition frequency between the TLS ground and excited states, $\mathbf{d}_{ge}$ is the dipole moment of the transition of the TLS, $c$ is the speed of light, $\hbar$ is the Planck constant, $\tau_P$ and $\tau_n$ are relaxation times of the polarization and the population inversion, respectively, and $n_0$ is the population inversion in the TLS in the absence of the field. The latter quantity characterizes the pump rate of incoherent radiation that creates the inversion.

We assume that the field and the polarization are slow functions of time with the carrier frequency $\omega_0$. Then, $E$ and $P$ can be represented in the form $E(x,t) = e(x,t)\exp(-i\omega_0 t) + e^*(x,t)\exp(i\omega_0 t)$ and $P(x,t) = p(x,t)\exp(-i\omega_0 t) + p^*(x,t)\exp(i\omega_0 t)$, where the complex-valued functions $e(x,t)$ and $p(x,t)$ are slowly changing during the oscillation period, $|\partial e(x,t)/\partial t| \ll \omega_0 |e(x,t)|$ and $|\partial p(x,t)/\partial t| \ll \omega_0 |p(x,t)|$. In this description, the dispersion of the TLS is taken into account by the explicit calculation of the layer dynamics. The dispersion of a metal is described by the Drude equation $\varepsilon(\omega) = 1 - \omega_p^2 (\omega^2 + i\gamma\omega)^{-1}$. Then, using the slow amplitude approximation we obtain[15]

$$\varepsilon(x)\frac{\partial^2 E(x,t)}{\partial t^2} = \left[ -i \left.\frac{\partial(\varepsilon(\omega,x)\omega^2)}{\partial \omega}\right|_{\omega_0} \frac{\partial e(x,t)}{\partial t} - \varepsilon(\omega,x)\omega_0^2 e(x,t) \right] e^{-i\omega_0 t}$$
$$+ \left[ i \left.\frac{\partial(\varepsilon(\omega,x)\omega^2)}{\partial \omega}\right|_{\omega_0} \frac{\partial e^*(x,t)}{\partial t} - \varepsilon(\omega,x)\omega_0^2 e^*(x,t) \right] e^{i\omega_0 t} \quad (4)$$

and

$$\frac{\partial(\varepsilon\omega^2)}{\partial\omega} = 2\omega \frac{(1-\mathrm{Re}\,\varepsilon)^2 + \mathrm{Re}\,\varepsilon(\mathrm{Im}\,\varepsilon)^2}{(\mathrm{Im}\,\varepsilon)^2 + (1-\mathrm{Re}\,\varepsilon)^2} + i\omega\,\mathrm{Im}\,\varepsilon\frac{(\mathrm{Im}\,\varepsilon)^2 - (1-\mathrm{Re}\,\varepsilon)^2}{(\mathrm{Im}\,\varepsilon)^2 + (1-\mathrm{Re}\,\varepsilon)^2} = 2\omega\bar{\alpha}(\omega,x). \quad (5)$$

Using Eqs. (4) and (5) we obtain Maxwell-Bloch equations for slow amplitudes

$$\frac{\partial^2 e(x,t)}{\partial x^2} + \frac{\omega_0^2}{c^2}\varepsilon(x)e(x,t) + i\frac{2\omega_0\alpha(x)}{c^2}\frac{\partial e(x,t)}{\partial t} = -4\pi\frac{\omega_0^2}{c^2}p(x,t), \quad (6)$$

$$\frac{\partial p(x,t)}{\partial t} + \frac{p(x,t)}{\tau_p} = -\frac{i\mu(x)^2 n(x,t) e(x,t)}{\hbar}, \quad (7)$$

$$\frac{\partial n(x,t)}{\partial t} + \frac{1}{\tau_n}(n(x,t) - n_0) = \frac{4}{\hbar}\mathrm{Im}\left(e^*(x,t)p(x,t)\right), \quad (8)$$

where $\alpha(x) = 0$ for dielectric layers.

### III. THE DEPENDENCE OF LASING THRESHOLD ON LOSSES (NUMERICAL SIMULATION)

For a finite photonic crystal system, Eqs, (6)-(8) can be solved by using the FDTD method. We consider the photonic crystal having 30 elementary cells of size $a = 200$ nm. In the crystal, the ratio of the widths of the dielectric and metallic layers is $d_d/d_m = 4$, and the transition frequency of the TLS is $\omega_0 = 10^{16}$ Hz. We also assume that the relaxation times of the polarization and the TLS population inversion are $\tau_p = 3 \cdot 10^{-14}$ sec and $\tau_n = 5 \cdot 10^{-12}$ sec, respectively, and the absolute value of the dipole moment is $|\mathbf{d}_{ge}| = 20$ dB.[16] Finally, we assume that the permittivity of dielectric layers is $\varepsilon_d = 3$ and the permittivity of the metal layers is the same as that of silver at the frequency of $\omega_0 = 10^{16}$ Hz ( $\text{Re}\,\varepsilon_m(\omega_0) = -1$ ). These parameters are typical for recently studied plasmon DFB lasers.[7-12]

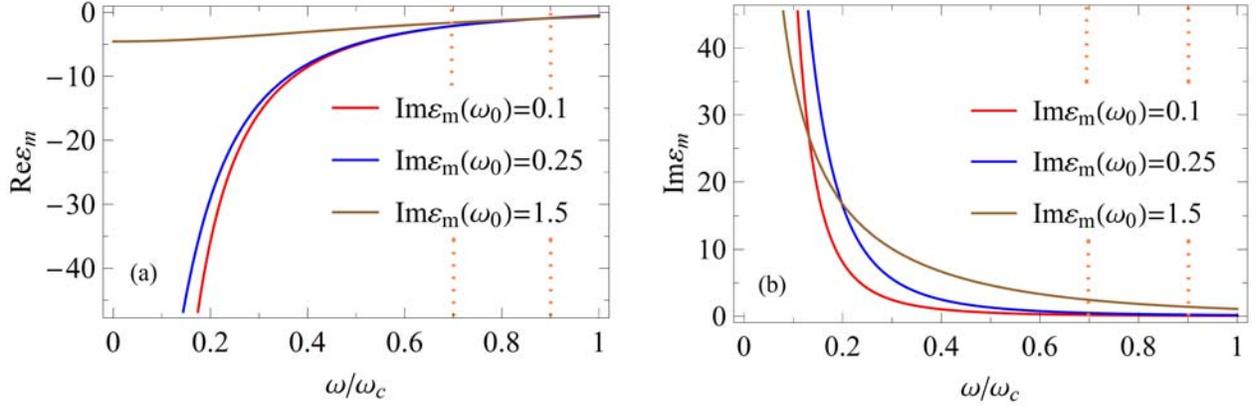

FIG. 2. The frequency dependence of the dielectric permittivity of the metal layers for different values of the imaginary part of the dielectric permittivity of these layers at the amplifying medium transition frequency, $\text{Im}\,\varepsilon_m(\omega_0)$. The frequency region of interest is between the vertical dotted lines. One can see from Fig. 2(b) that in this region, an increase of $\text{Im}\,\varepsilon_m(\omega_0)$ is accompanied by an increase of $\text{Im}\,\varepsilon_m(\omega)$.

The chosen transition frequency of the TLS falls into the gap of the photonic crystal, $0.985 \times 10^{16}\,\text{Hz} < \omega < 1.3 \times 10^{16}\,\text{Hz}$. Therefore, the lasing mode is a standing wave of the gap edge. Thus, there is a detuning $\Delta = 1.5 \times 10^{14}\,\text{Hz}$ between the TLS transition frequency and the frequency of the resonator mode. For the chosen values of the system parameters, the dielectric permittivity of the metal layers changes only weakly near the center of the second band gap, $\omega_c = 1.14 \times 10^{16}\,\text{Hz}$, in the range of frequencies of $0.7\omega_c < \omega < 0.9\omega_c$ (see Fig. 2).

In numerical simulations, for initial conditions we choose a random distribution of the field in the photonic crystal in the absence of the polarization and the population inversion in the active medium. Then we look for the steady state of field generation.

In the laser, the dependence of the energy on the pump rate, shown in Fig. 3, has a standard form. At the threshold, the energy of the field starts increasing with an increase of the population inversion. When loss increases, the slope of the generation line decreases.[1] Above the threshold, as shown in Sec. IV, the generated energy is inversely proportional to $\text{Im}\,\varepsilon_m$. This explains the decrease of line slopes in Fig. 3 with a loss increase.

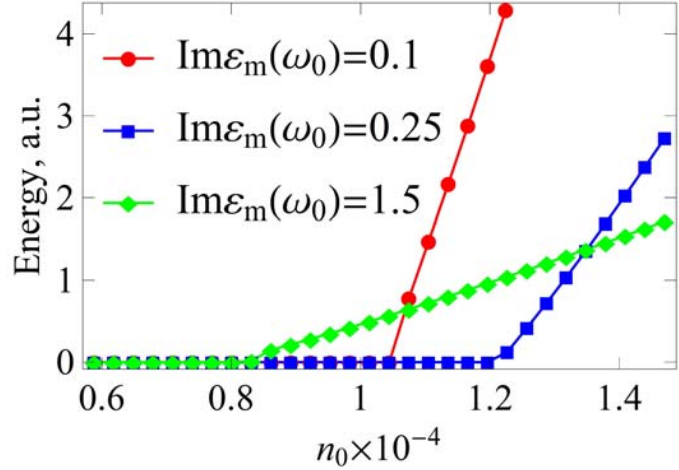

FIG. 3. Generation curves (the energy of the electromagnetic field summed over the photonic crystal) calculated for different values of the imaginary part of the dielectric constant of the metal.

As Fig. 4 shows, the dependence of the population inversion threshold non-monotonically depends on the loss in metal layers. For small losses, its increase results in an increase of the threshold. Then, above some critical value of the loss, its increase leads to a decrease of the population inversion threshold, which starts increasing for further loss increase.

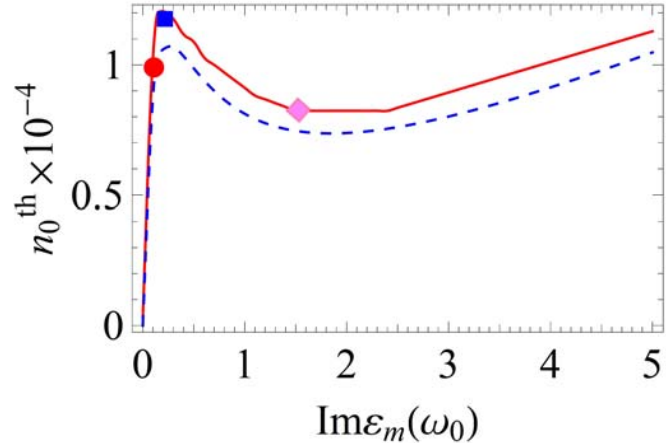

FIG. 4. The dependence of the population inversion threshold on the imaginary part of the dielectric permittivity of the metal at the transition frequency of the amplifying medium. The solid line is numerical simulation. The dashed line is plotted in accordance with the theory developed in Sec. IV.

In obtaining the curves in Fig. 4, the numerical simulations have been done for a finite photonic crystal containing 30 elementary cells while analytical calculations have been conducted for an infinite crystal. There is additional radiation from the ends of the finite crystal that is absent in the infinite crystal. Additional radiation results in an increase of the generation threshold that explains some difference between the numerical and analytical results that can be seen in Fig. 4.

In Fig. 5, the dependence of the TLS lasing frequency $\omega_g$ on loss in the metal is shown. When loss increases, $\omega_g$ is pulled towards the transition frequency of the TLS. As we mention above, the latter is positioned in the photonic bandgap of the crystal. Below we show that pulling the lasing frequency towards the transition frequency of the TLS results in a decrease of the lasing threshold.

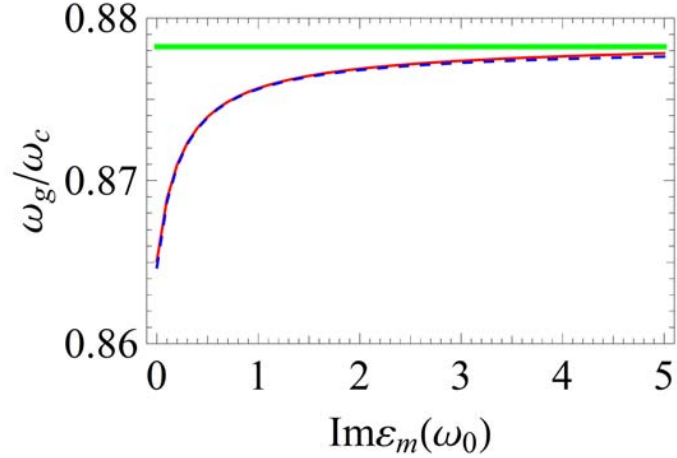

FIG. 5. The lasing frequency $\omega_g$ near the threshold as a function of loss in the metal layers (in units $\omega_c = c/a$, which is the frequency at the center of the second band gap). The results of numerical simulations and analytical calculations are shown by solid red and blue dashed lines, respectively. The TLS transition frequency, $\omega_0$, is shown by the solid green line.

## IV.   THE DEPENDENCE OF LASING THRESHOLD ON LOSSES (ANALYTICAL APPROACH)

In this section, we obtain the dependence of the threshold population inversion $n_0^{th}$ on the imaginary part of the dielectric permittivity of metal, $\text{Im}\,\varepsilon_m$, for an infinite photonic crystal. As it follows from numerical simulations, in the laser under study, the regime of a single mode lasing is realized (Fig. 5). In this regime, the field distribution of an infinite photonic crystal coincides with the distribution of one of the modes described by the Helmholtz equation[15]

$$\frac{\partial^2 E_M(x)}{\partial x^2} + \text{Re}[\varepsilon(\omega_M,x)]\frac{\omega_M^2}{c^2} E_M(x) = 0, \qquad (9)$$

with the periodic boundary conditions. Here $\varepsilon(\omega_M, x)$ is either $\varepsilon_d$ or $\varepsilon_m(\omega_M)$ in dielectric or metal layers, respectively. $E_M(x)$ and $\omega_M$ are the field distribution and the frequency of the mode. In this approach, Eq. (1) is split into two equations. The first one, Eq. (9), defines the special distribution of the lasing mode. The second equation, Eq. (10) below, gives the time evolution of the amplitude of this mode. The numerical analysis shows that the imaginary part of the value of permittivity of metal and the inversion population do not substantially affect the field distribution in the system (Fig. 6). Thus, compared to Eq. (1), in Eq. (9) the imaginary part of the dielectric permittivity of metal and the temporal dependence of the active medium controlled by the population inversion are excluded because they affect the field distribution in the lasing mode only weakly.[17]

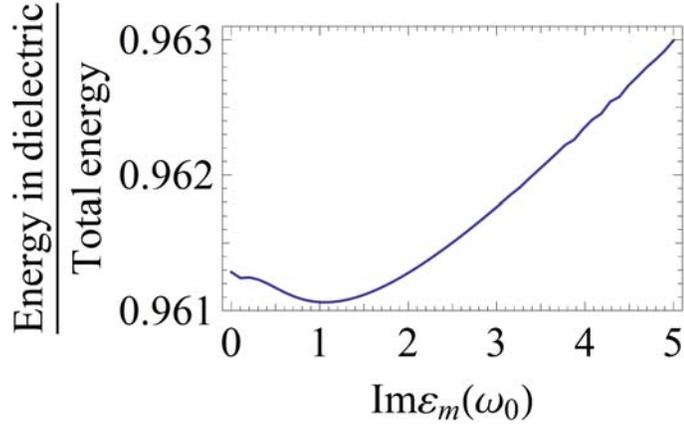

FIG. 6. The ratio of the energies localized in the dielectric and the total energy in the photonic crystal. The pump rate is fixed, $n_0 = 1.5 \times 10^{-4}$.

To determine the lasing threshold, it is sufficient to solve a linear problem neglecting the dependence of the population inversion $n$ on the field amplitude.[14] In this case, $n = n_0$ and it does not depend on time. This allows for a factorization of the generated field, $E(x,t) = E_M(x)e(t)$, after which Eq. (1) is simplified as

$$-\text{Re}[\varepsilon(x)]\omega_M^2 E_M(x)e(t) - \varepsilon(x)E_M(x)\frac{\partial^2 e(t)}{\partial t^2} = 4\pi \frac{\partial^2 P(x,t)}{\partial t^2}, \qquad (10)$$

After multiplying Eq. (10) by $E_M^*(x)$ and integrating it over the whole space we obtain

$$-U\omega_M^2 e(t) - (U + i\Gamma)\frac{\partial^2 e(t)}{\partial t^2} = \frac{1}{2}\frac{\partial^2 p(t)}{\partial t^2}, \qquad (11)$$

where

$$\Gamma = \frac{1}{8\pi} \int \text{Im}[\varepsilon(x)] E_M(x) E_M^*(x) dx,$$

$$U = \frac{1}{8\pi} \int \text{Re}[\varepsilon(x)] E_M(x) E_M^*(x) dx,$$

$$p = \int E_M(x) P(t,x) dx.$$

Similarly, Eq. (2) can be reduced to

$$\frac{\partial^2 p(t)}{\partial t^2} + \frac{2}{\tau_p} \frac{\partial p(t)}{\partial t} + \omega_0^2 p(t) = -8\pi i \omega_0 \mu n_0 e(t)/\hbar, \qquad (12)$$

where

$$\mu = \int |d_{eg}|^2 E_M(x) E_M^*(x) dx.$$

From Eqs. (12) and (13) we can find $n_0^{th}$ that characterizes the threshold pump rate. At the onset of lasing, time dependencies of the electric field and the polarization have the form $e(t) = e_0 \exp(-i\omega_g t)$ and $p(t) = p_0 \exp(-i\omega_g t)$, where $e_0$ and $p_0$ are constants and $\omega_g$ is the lasing frequency that is determined by the condition of the onset of lasing. As a result, we obtain

$$-U\omega_M^2 e_0 + (U + i\Gamma)\omega_g^2 e_0 = -\frac{1}{2}\omega_g^2 p_0, \qquad (13)$$

$$-\omega_g^2 p_0 - \frac{2}{\tau_p} i\omega_g p_0 + \omega_0^2 p_0 = -16\pi \omega_0 \mu n_0 e_0 /\hbar. \qquad (14)$$

Since the lasing mode frequency and the frequency of the TLS transition are close, we obtain the final equations

$$\left[\frac{\Gamma \omega_{MP}}{2U} - i(\omega_g - \omega_M)\right] e_0 = \frac{i\omega_0}{4U} p_0, \qquad (15)$$

$$\left[\frac{2}{\tau_p} - i(\omega_g - \omega_0)\right] p_0 = -i8\pi \mu n_0 \hbar e_0. \qquad (16)$$

Parameters $\omega_g$ and $n_0$, for which the nontrivial solution of Eqs. (15) and (16) arises, determine the lasing frequency and the threshold population inversion:

$$\omega_g = \frac{\Gamma \omega_M \tau_p \omega_0 + 2U\omega_M}{\Gamma \omega_M \tau_p + 2U}, \qquad (17)$$

$$n_0^{th} = \frac{\hbar \Gamma}{4\pi\mu\tau_p}\left[1+(\omega_0-\omega_M)^2\left(\frac{\Gamma\omega_M}{2U}+\frac{1}{\tau_p}\right)^{-2}\right]. \qquad (18)$$

If we identify $2U/\Gamma\omega_M$ as the longitudinal relaxation time of the laser, $\tau_a$, and $\sqrt{2\pi\mu\omega_M/\hbar U}$ as the interaction constant of the field with the amplifying medium, $\Omega_R$, then Eqs. (17) and (18) coincide with well-known equations for $\omega_g$ and $n_0^{th}$:[1]

$$\omega_g = \frac{\tau_p\omega_0+\tau_a\omega_M}{\tau_p+\tau_a}, \qquad (19)$$

$$n_0^{th} = \frac{1}{\Omega_R^2\tau_p\tau_a}\left[1+(\omega_0-\omega_M)^2\left(\tau_a^{-1}+\tau_p^{-1}\right)^{-2}\right]. \qquad (20)$$

Thus, the generation curves, shown in Fig. 2 are the same as for a laser with the uniform distribution of the field in the resonator. These curves are described by the equation[1]

$$|e_{st}|^2 = \tau_a\left(n_0-n_0^{th}\right)/4\tau_n, \qquad (21)$$

where $e_{st}$ is the stationary field amplitude above the threshold. Note that a non-monotonic behavior of the threshold population inversion described by Eq. (20) is a feature of single mode lasers. In a multimode laser with the Fabry-Perot resonator, modes are very close to each other so that there is practically no frequency detuning.

The results obtained in this section allow one to explain the non-monotonic dependence of the threshold of the population inversion described by Eq. (18) and shown in Fig. 3. The threshold increases linearly with respect to $\Gamma$ for small and large losses

$$\begin{aligned}n_0^{th} &= \frac{\hbar\Gamma}{4\pi\mu\tau_p}\left[1+(\omega_0-\omega_M)^2\tau_p^2\right], \quad \Gamma \ll 2U/\omega_M\tau_p, \\ n_0^{th} &= \frac{\hbar\Gamma\omega_M}{4\pi\mu\tau_p\omega_0}, \quad \Gamma \gg 2U|\omega_0-\omega_M|/\omega_M.\end{aligned} \qquad (22)$$

$n_0^{th}$ has a minimum for an intermediate value of $\Gamma$. The non-monotonic behavior with loss increase is due to an increase of broadening of the lasing mode line. This broadening results in a shift of the lasing frequency towards the transition frequency of the amplifying medium (Fig. 5). The interaction between the electromagnetic wave and the amplifying medium peaks when these two frequencies coincide. It decreases when the frequencies move away from each other. As a result, the loss increase in the resonator causes an increase of the attenuation rate in the metallic layers and a field increase in the amplifying layers. Depending on which of these competing factors prevails, the lasing threshold may either increase or decrease with the loss increase in the resonator. In Fig. 4, for $\operatorname{Im}\varepsilon_m < 2U\left[\tau_\sigma \omega_M \int_{metal} |E(x)|^2 dx\right]^{-1} \approx 0.2$ and $\operatorname{Im}\varepsilon_m > 2U|\omega_0 - \omega_M|\left[\omega_M \int_{metal} |E(x)|^2 dx\right]^{-1} \approx 3$ the lasing threshold is mainly defined by the energy loss in the metal, while for $2U\left[\tau_\sigma \omega_M \int_{metal} |E(x)|^2 dx\right]^{-1} < \operatorname{Im}\varepsilon_m < 2U\left[\tau_\sigma \omega_M \int_{metal} |E(x)|^2 dx\right]^{-1}$, an increase of the interaction between the amplifying medium and the electromagnetic field plays the main role. This increase can be described by considering an overlap of the resonator mode line and the transition line of the amplifying medium (Fig. 7). When loss in the resonator is small ($\operatorname{Im}\varepsilon_m < 2$), the overlap increases with the loss increase. Then, for $\operatorname{Im}\varepsilon_m \leq 2$, it decreases. The largest overlap and the lowest generation threshold occur for the same value of $\operatorname{Im}\varepsilon_m$ (Fig. 8).

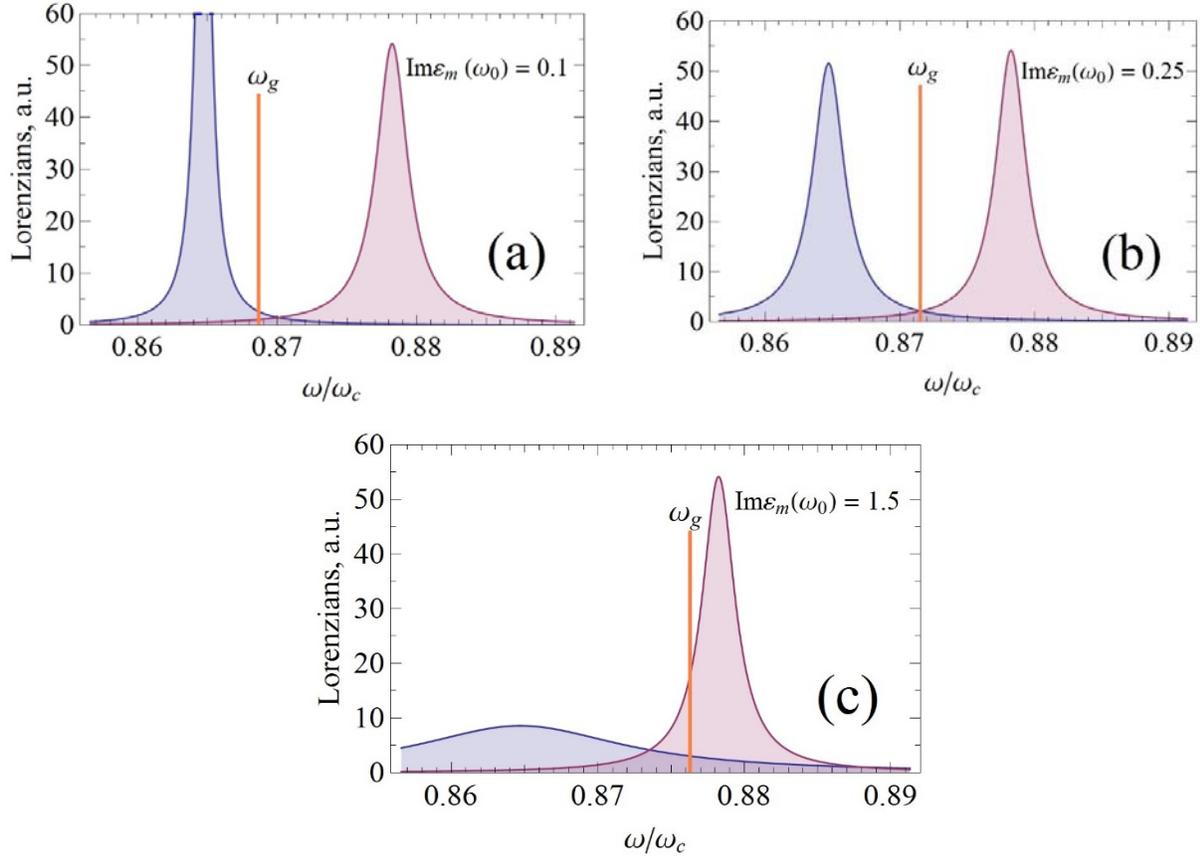

FIG. 7. Lorentzians of the resonator and transition lines of the amplifying medium for values (a) $\operatorname{Im}\varepsilon_m = 0.1$, (b) $\operatorname{Im}\varepsilon_m = 0.25$, and (c) $\operatorname{Im}\varepsilon_m = 1.5$. The Lorentzians are normalized so that the area under each curve is equal to unity. Blue and red lines correspond to the resonator and amplifying medium lines, respectively. The orange line shows the lasing frequency. The pumping frequency is $\omega_0 = 0.878\omega_c$ and the frequency of the resonator mode is $\omega_M = 0.864\omega_c$.

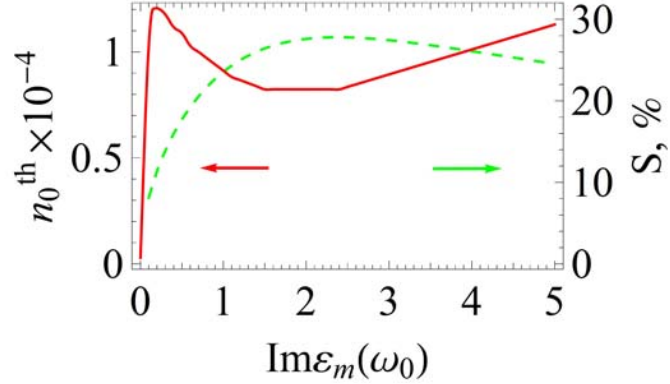

FIG. 8. Dependencies on $\text{Im}\,\varepsilon_m$ of the lasing threshold (the red solid line) and the ratio of areas of Lorentzians of the amplifying medium and the resonator mode overlap (the green dashed line).

## V. DISCUSSIONS AND CONCLUSIONS

Since a DFB laser works at the boundary of the bandgap,[12, 18] thanks to the Borrmann effect,[19] the field is mainly concentrated in active dielectric layers. Hence, the field distribution in the system changes only slightly when the loss in the metal increases. As one can see from Fig. 9, the fraction of the generated energy localized in the metal changes only by about 5% while the imaginary part of the metal permittivity increases by the factor of 5. As shown in Sec. IV, in the absence of frequency detuning, the threshold value of the population inversion that is needed for laser generation is

$$n_0^{th} \sim \text{Im}\,\varepsilon_m \times I_{md}, \qquad (23)$$

where $I_{md} = \int_{metal} |E_M(x)|^2 dx \Big/ \int_{dielectric} |E_M(x)|^2 dx$, $E_M(x)$ is the electric field distribution in the lasing mode.

As one can see from Eq. (23), when the fraction of the generated energy localized in the metal changes by 5% while $\text{Im}\,\varepsilon_m$ increases from 0.4 to 1.5, the threshold population inversion $n_0^{th}$ triples. This is also confirmed by numerical simulations. In the presence of detuning, $n_0^{th}$ is no longer described by Eq. (23). It is now described by Eq. (18) that gives the dependence shown in Fig. 3. In this case, when $\text{Im}\,\varepsilon_m$ increases from 0.4 to 1.0, $I_{md}$ and $n_0^{th}$ are moving in opposite directions; the former increases and the latter decreases. Thus, as opposed to Refs. 3, 4, the non-monotonic behavior of the lasing threshold is not caused by changes in the spatial overlap of the lasing mode and the active medium layers.

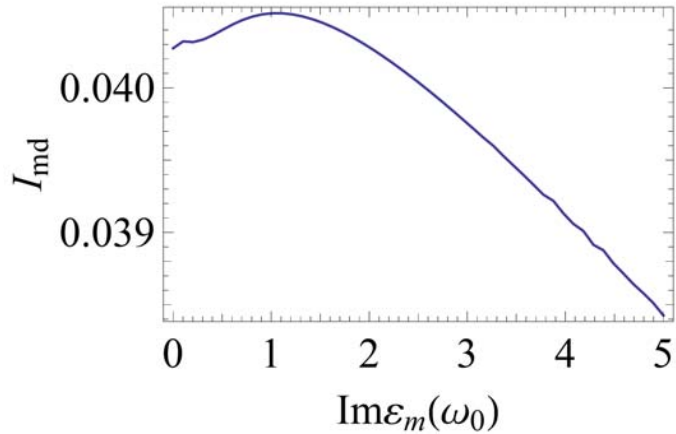

FIG. 9. The ratio of the energies localized in metal and dielectric in the photonic crystal. The pump rate is fixed at $n_0 = 1.5 \times 10^{-4}$.

To conclude, we show that in a DFB laser in which the transition frequency of the amplifying medium is detuned from the frequency of the resonator mode, an increase in the loss may lead to a decrease in the lasing threshold. Such a decrease is due to the broadening of the resonator mode that leads to the lasing frequency being pulled towards the transition frequency of the amplifying medium and to an increase of the overlap between the lines of the resonator mode and the transition of the amplifying medium. As a result, the interaction between an electromagnetic wave and the amplifying medium also increases. Thus, while the loss increases the dissipation rate in the resonator, it nonetheless results in an increase of the amplification rate in the active medium. If the latter factor prevails, the conditions for lasing are improved.

Our results allow for selecting the parameters of a DFB laser for which an increase of the imaginary part of the dielectric permittivity is not critical. Moreover, such an increase may improve laser characteristics. This is important in connection with the ongoing development of plasmonic DFB lasers.[7-12]

### ACKNOWLEDGEMENTS

The work was supported by RFBR grants No. 13-02-92660, by Dynasty Foundation, and by the NSF under Grant No. DMR-1312707.